\documentclass[useAMS,usenatbib]{mn2e}
\usepackage{times}

\usepackage{graphicx}

\title[Warp/Canis Major]{Galactic Warp in the overdensity
   of the Canis Major Region}
\author[M. L\'opez-Corredoira]{M. L\'opez-Corredoira$^{1}$\thanks{E-mail:
martinlc@iac.es} \\
$^{1}$ Instituto de Astrof\'\i sica de Canarias, C/.V\'\i a L\'actea, s/n,
E-38200 La Laguna (S/C de Tenerife), Spain}
\begin{document}

\date{Accepted xxxx xxx xx. Received xxxx xxx xx; in original form xxxx xxx xx}

\pagerange{\pageref{firstpage}--\pageref{lastpage}} \pubyear{2005}

\maketitle

\label{firstpage}

\begin{abstract}
Bellazzini et al.\ (2006b) claim that L\'opez-Corredoira et al.'s (2002)
warp model is totally unable to reproduce the Canis Major structure in 
the red clump stars. However, slight variations in the azimuth of 
the L\'opez-Corredoira et al.\ (2002) warp model, 
justified by the uncertainties in the parameter as well as
the local irregularities with respect to the average model, derive a result 
much closer to the observations of the overdensity south/north. 
The bump of red clump stars with $m_K=13$--13.5
around $l=241^\circ $, $b=-8.5^\circ $ and the depth of the
Canis Major structure are also explainable in terms of 
the warp 
with an appropriate extrapolation 
of constant height between galactocentric radii of 13 and 16 kpc, 
as observed roughly in the southern warp,
instead of a monotonically increasing height like the northern warp;
and the observed velocity distribution of stars
cannot exclude the warp possibility.
A warp model is therefore still a possible
explanation of the Canis Major overdensity, and the hypothesis 
of the existence of a dwarf galaxy is unnecessary, although 
still a possibility too.
\end{abstract}


\section{Introduction}

Bellazzini et al.\ (2006b, hereafter B06b)\footnote{
I was the referee
of this paper for MNRAS and accepted it for publication after a pair
of revisions because I considered that the authors had done an
interesting analysis which should be known, even though I did not agree
with some points. I had recommended the authors to analyze some weak
points but they did not follow my advice, so I perform here the
analysis of these weak points (\S 2.1 and 2.2) and some
others (\S 2.3 and 2.4) which I thought of after the acceptance
of Bellazzini et al. (2006b).} 
have claimed that the
excess of red clump stars in southern with respect to northern galactic 
latitudes between  $l=200^\circ$ and 280$^\circ $ is most probably 
associated with a new dwarf galaxy, namely the Canis Major (CMa) galaxy, 
so that it is far from being explained in terms of the known characteristics 
of the Galactic warp model derived from the parameters given by 
L\'opez-Corredoira et al.\ (2002, hereafter L02); they conclude that the warp
{\it ``is totally unable to reproduce the CMa structure''}.
Previous papers (e.g.\ Martin et al.\ 2004, Mart\'\i nez-Delgado
et al.\ 2005a) have presented other proofs in favour of the existence of the 
dwarf galaxy and against the warp possibility. 
The discussion on the validity of the proofs is a topic of heated debate 
nowadays, with authors like Momany et al.\ 
(2004, 2006[hereafter M06]) arguing that these proofs can 
also be reproduced by the Galactic warp+flare.

Here, I also want to contribute to the debate showing 
that slight variations in L02 warp model 
derive a result much closer to the observations of the overdensity
(\S \ref{.overdensity}).
The bump of red clump stars with $m_K=13$--13.5
around $l=241^\circ $, $b=-8.5^\circ $ (\S 2.2) 
and the depth of the Canis Major structure (\S 2.3) 
are also explainable in terms of 
the warp. I shall
make a few comments about the velocity distribution of its stars (\S 2.4). 

\

\section{Galactic warp vs.\ dwarf galaxy hypothesis} 

\subsection{Canis Major overdensity}
\label{.overdensity}

\begin{figure}
{\par\centering \resizebox*{8.4cm}{7cm}{\includegraphics{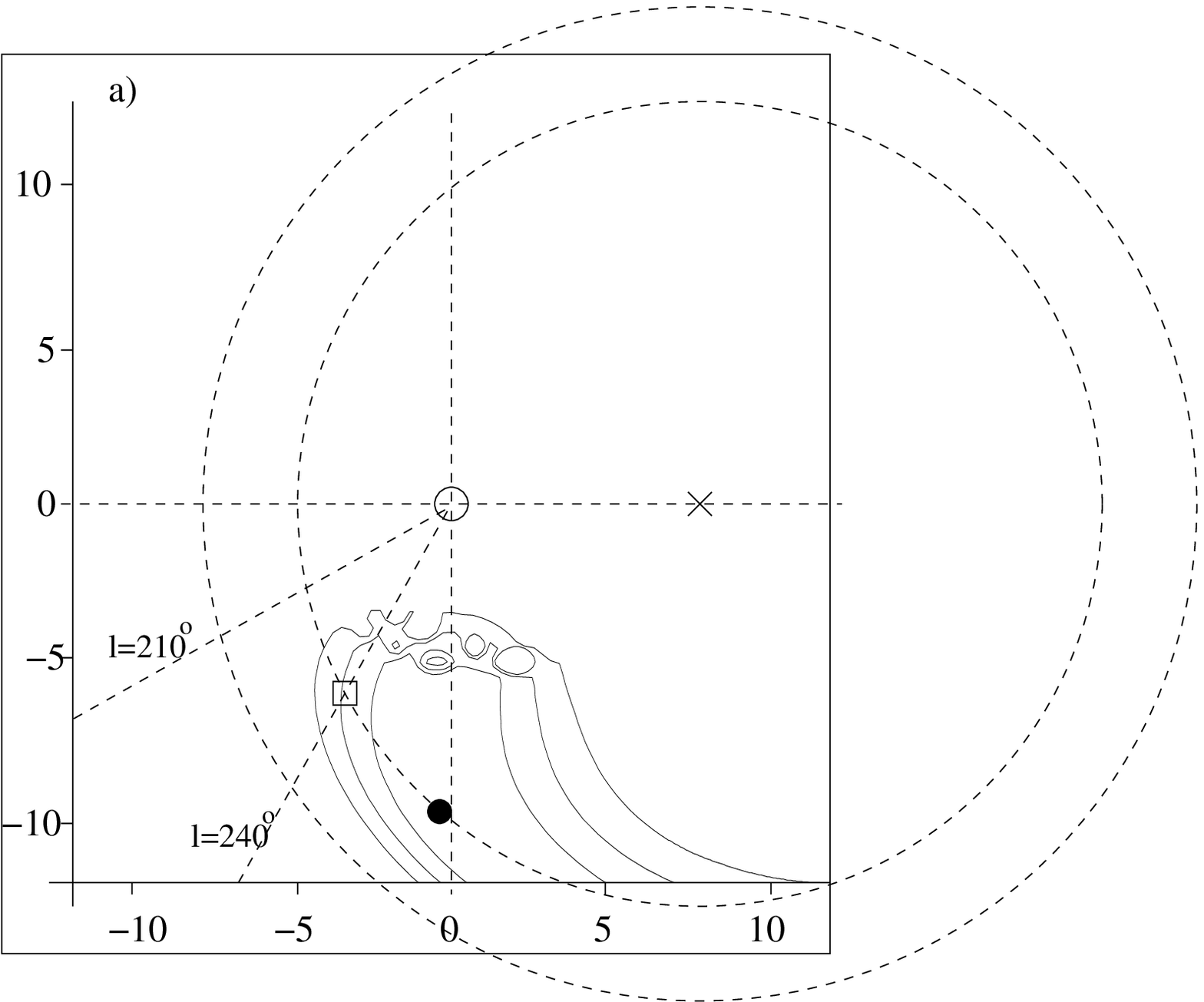}}
\hspace{-2.6cm}\resizebox*{8.4cm}{7cm}{\includegraphics{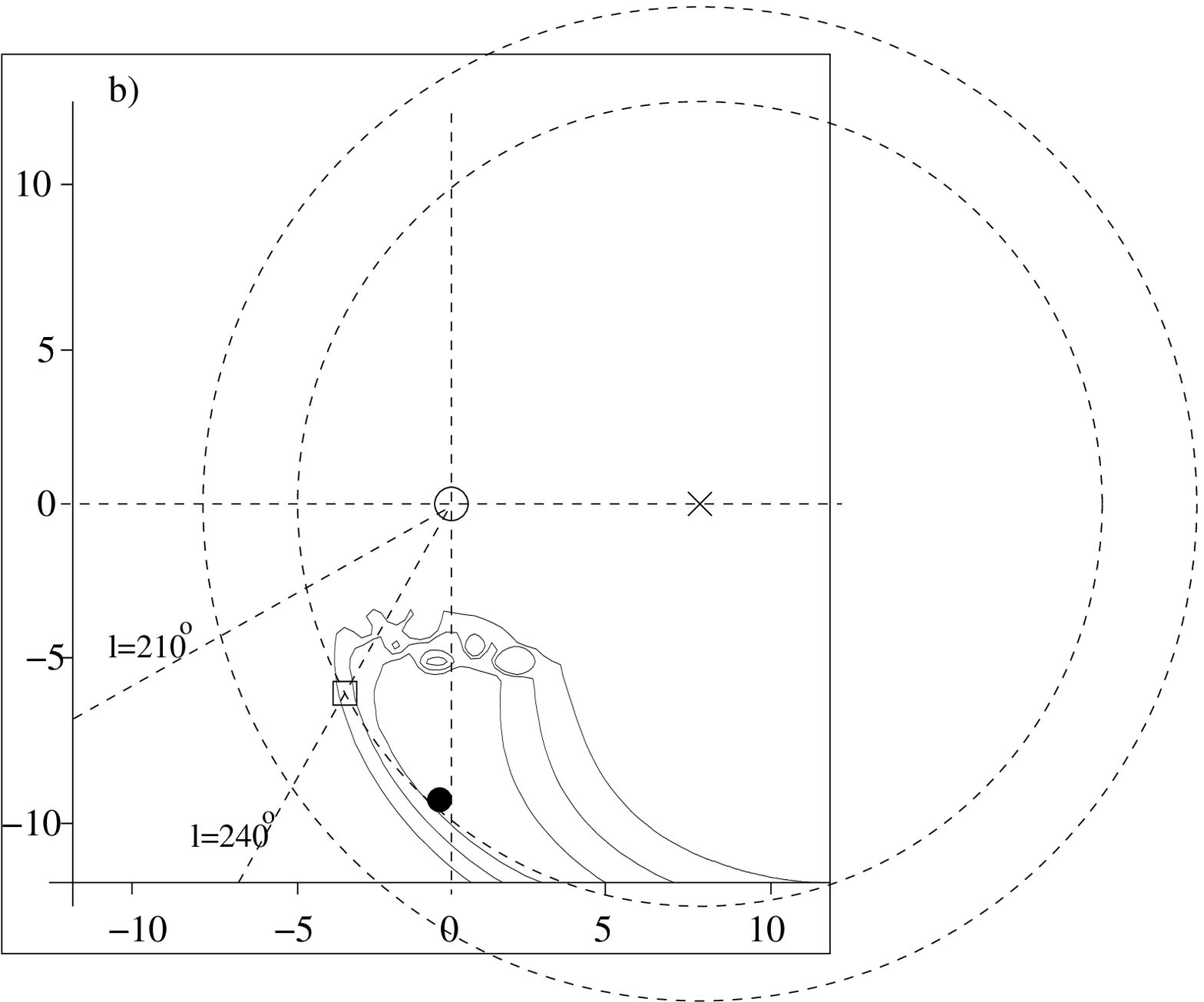}}\par}
\vspace{-0.7cm}
{\par\centering \resizebox*{8.4cm}{7cm}{\includegraphics{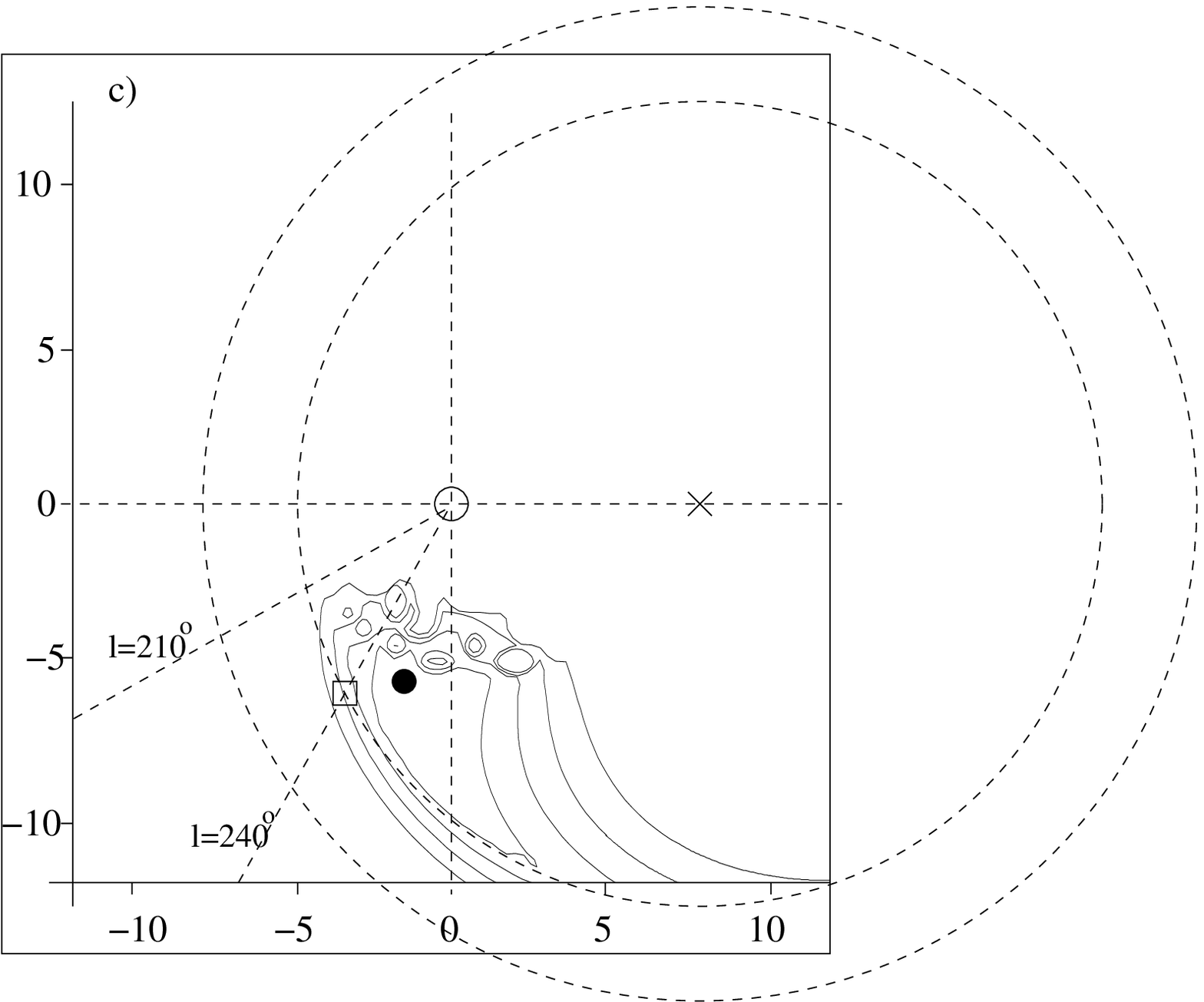}}
\hspace{-2.6cm}\resizebox*{8.4cm}{7cm}{\includegraphics{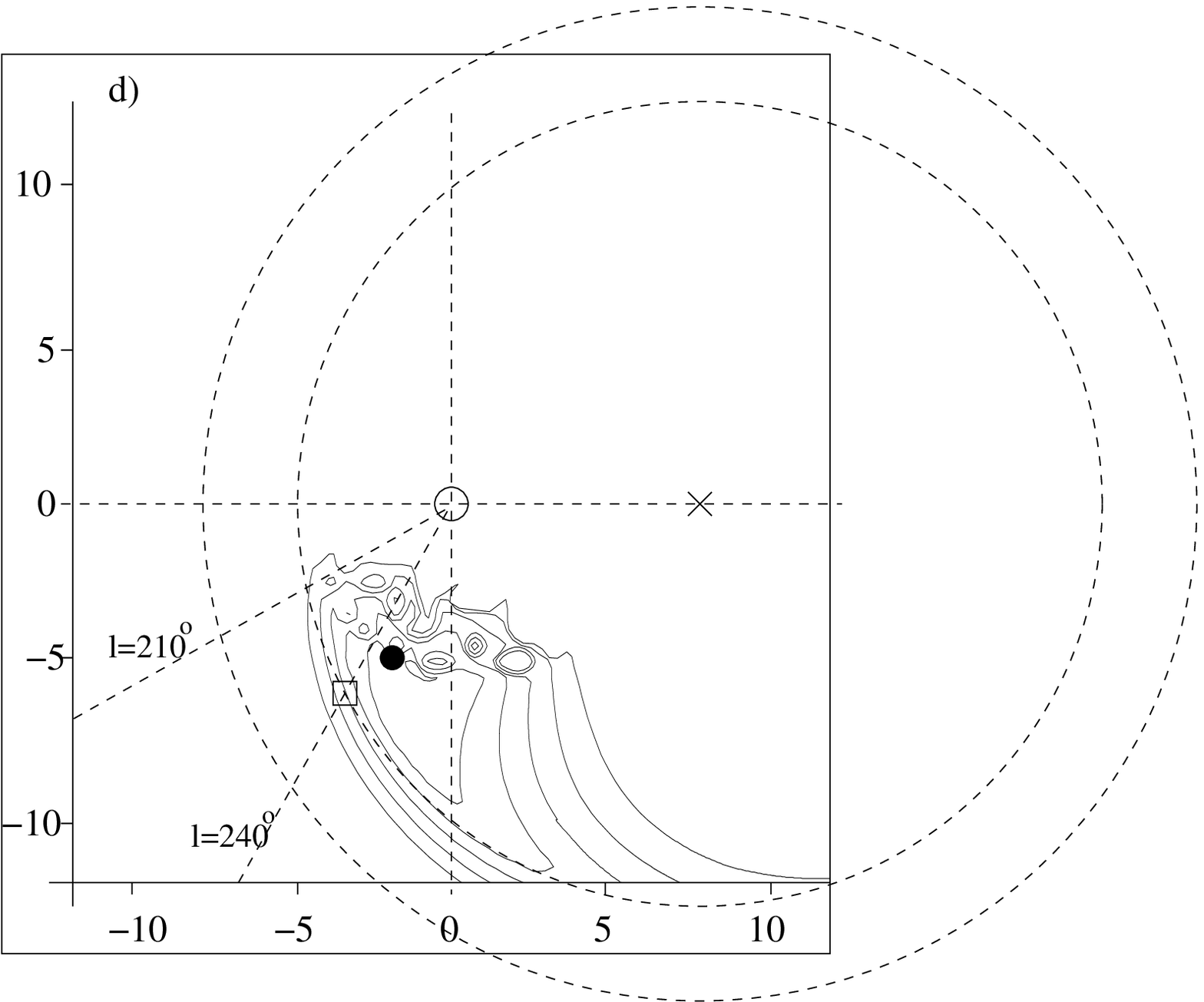}}\par}
\vspace{-0.7cm}
{\par\centering \resizebox*{8.4cm}{7cm}{\includegraphics{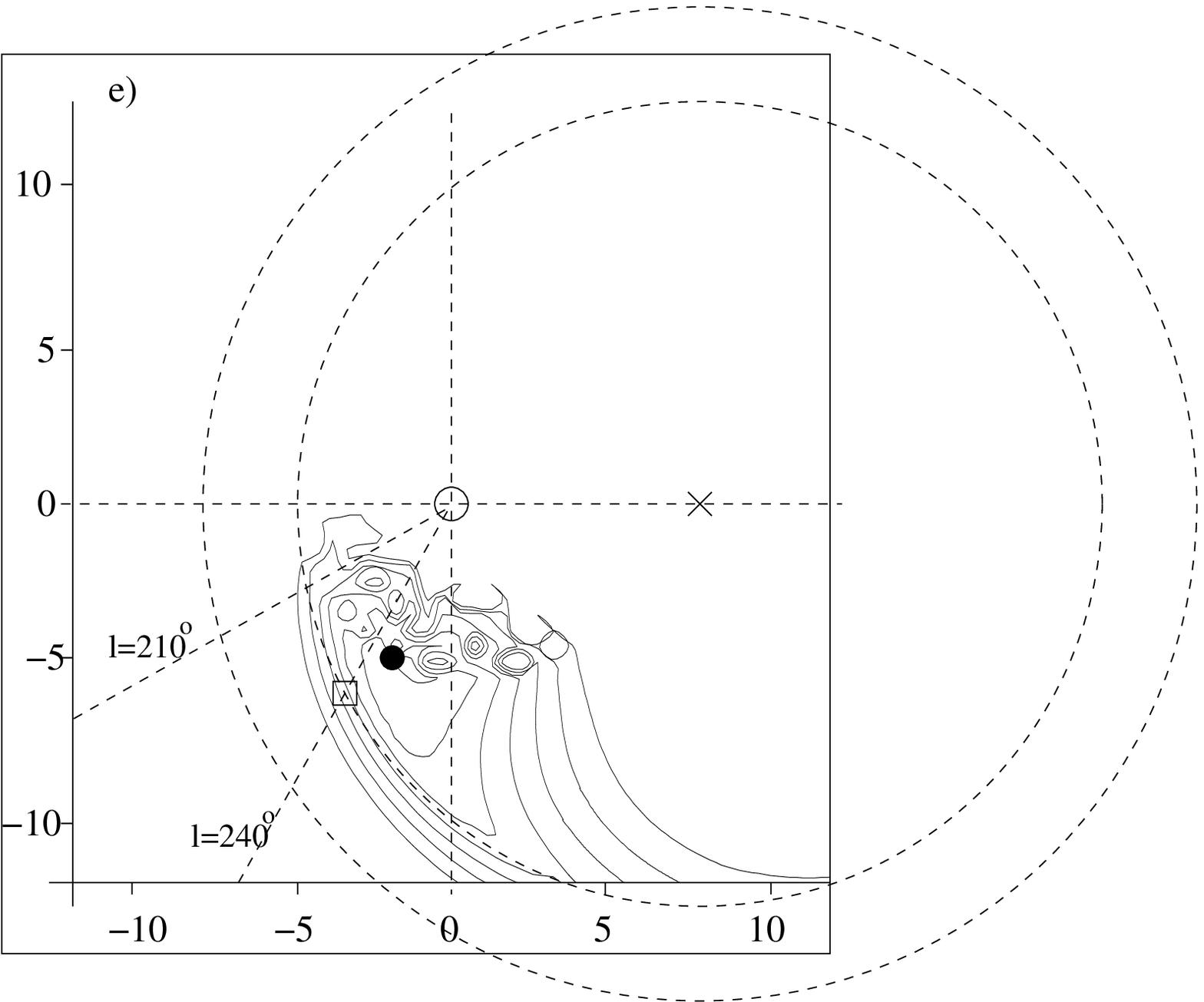}}
\hspace{-2.6cm}\resizebox*{8.4cm}{7cm}{\includegraphics{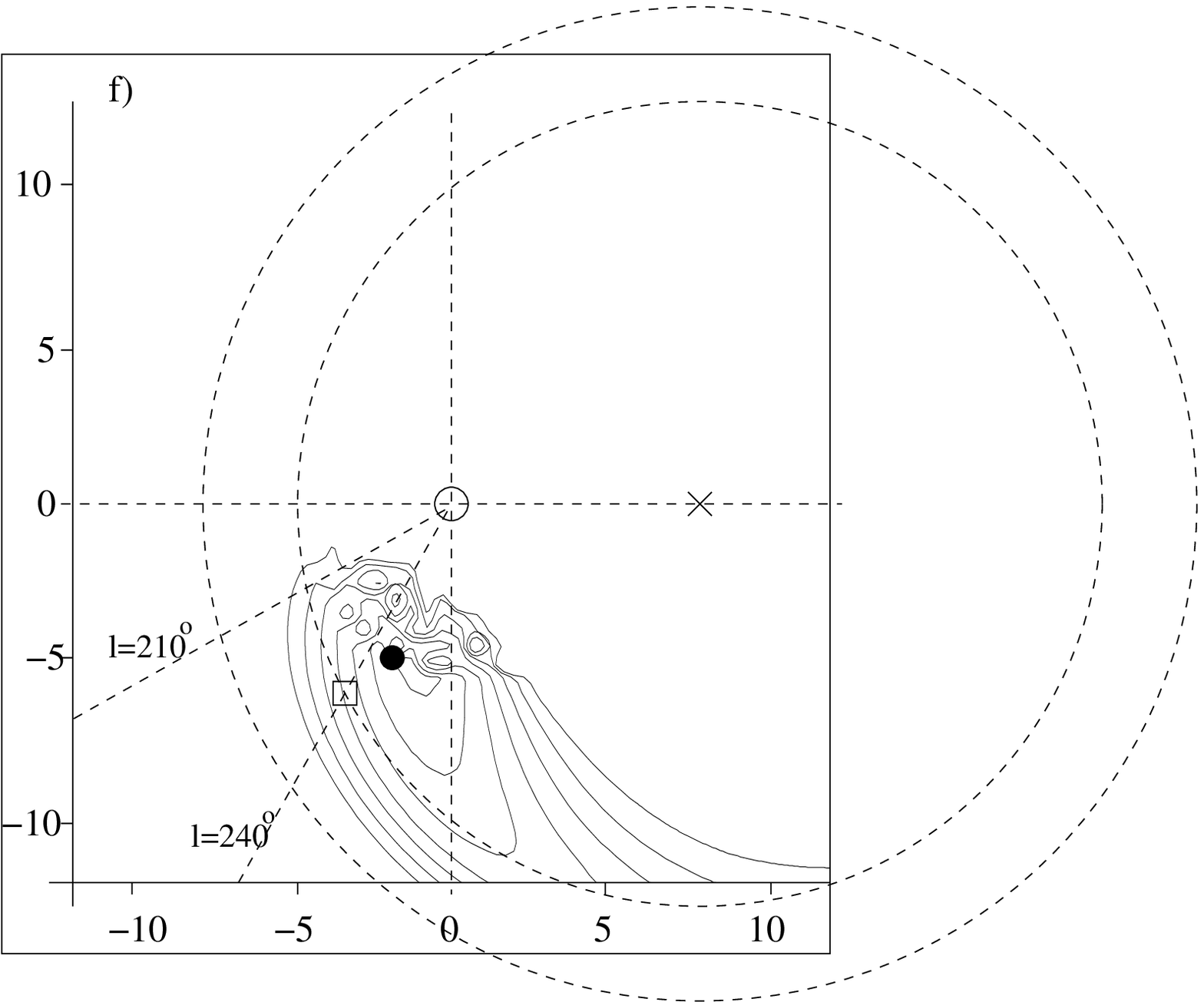}}\par}
\caption{a) Subtracted density maps
(first contour: 1000 star/kpc$^2$ and following contours
every 500 star/kpc$^2$; length units kpc; Sun at the center, the cross points out the Galactic center,
the square indicates the maximum of the density map in B06b data;
the filled circle indicates tha maximum of the present density map;
the dashed circles mark the regions at distances 13 and 16 kpc respectively
from the centre of the Galaxy) 
calculated in the same way as B06b:
$\rho (S)-1.2\rho (N)$ integrated 
over $5^\circ<|b|<15^\circ $, excluding $|z|<0.5$ kpc;
with the warp model by L02 ($R_\odot=8$ kpc) for $R\le 16$ kpc
and no warp beyond 16 kpc. 
Maximum at $l=267^\circ$, distance from the Sun: 10.0 kpc.
b) Same as a), but with constant
height of the warp $z_w$ for $13<R<16$ kpc, and null beyond.
Maximum at $l=267^\circ$, distance from the Sun: 9.5 kpc.
c) Same as b), but with $\phi _W=0$
as azimuth of the warp. Maximum at $l=255^\circ$, 
distance from the Sun: 5.7 kpc.
d) Same as b), but with $\phi _W=+5^\circ$
as azimuth of the warp. Maximum at $l=248^\circ$, 
distance from the Sun: 5.4 kpc.
e) Same as b), but with $\phi _W=+10^\circ$
as azimuth of the warp. Maximum at $l=248^\circ$, distance from the Sun: 
5.4 kpc. f) Same as b), but with $\phi _W=+5^\circ$
as azimuth of the warp, and the Drimmel \& Spergel (2001) formula for warp 
amplitude. Maximum at $l=248^\circ$, distance from the Sun: 5.4 kpc.}
\label{Fig:simul}
\end{figure}

The height of the L02 warp over the plane defined by the central disc 
as a function of the galactocentric distance and azimuth is

\begin{equation}
z_w(R,\phi )[pc]=C_WR({\rm kpc})^{\epsilon _W}\sin [\phi -\phi_w]
\label{warp1}
,\end{equation}
with $\epsilon _W=5.25\pm 0.5$,
$\phi _W=-5\pm 5^\circ $ and $C_W=1.2\times 10^{-3}$ (for the given 
values of $\epsilon _W=5.25$ and $\phi _W=-5^\circ $; the normalization
will change if $\epsilon _W$ and/or $\phi _w$ vary).
B06b calculate in their fig.\ 6(down) the south/north overdensity of the 
red clump stars,
$\rho (S)-1.2\rho (N)$ due to this L02 warp
and the flared disc parameters given in L02\footnote{There is an error
in L02: ``+15'' pc was written in eq. (20)  instead of ``$-$15''; the correct value is
$-$15 pc because the Sun is ``above'' the plane. The calculations in L02
were correctly carried out with $-$15, and the error is just 
typographical. In any case, this does not affect too much to the
analysis of the warp at high values of $R$.}. The factor 1.2 was put by B06b to 
compensate for the average south/north asymmetry.
The asymmetry between
south and north is not  constant, and it is a very bad approximation
to take it as B06b does, but I use their expression in order to show that
I can roughly reproduce the results of their fig.\ 6 with the warp.
We confirm their results in Figure\ \ref{Fig:simul}a).
Effectively, the maximum of the overdensity is around 270$^\circ $
as they claim (and contrary to the claim by M06, who used a
different maximum definition and different constraints on region
selection), somewhat far from the centre of
the observed CMa structure at $l\approx 244^\circ $ (B06b, Fig. 6, top).

L02 only give a warp formula for $R\le 13$ kpc; beyond 13 kpc,
different extrapolations are possible.
We also make the same plot with a different extrapolation of the L02
model over the range $13<R(kpc)<16$, with a constant height of the
warp, $z_w(13\ {\rm kpc}<R<16\ {\rm kpc})=z_w(13\ {\rm kpc})$, 
which is more similar to the real gas 
southern warp (see \S 2.2). 
This is shown in Fig.\ \ref{Fig:simul}b); the result does not change too much.
As noted by B06b, the extrapolation beyond 13 kpc is not too important
because the CMa feature occurs approximately at this galactocentric
distance.

However, some attention should be paid to other parameters of the
warp. L02 have given an approximate model of the ``average'' warp for
the whole sky assuming  north--south symmetry and a power law for the
amplitude of this warp. This assumption is just a first-order approximation
since, as  is well known, the Galactic warp is not symmetric (Burton 1988;
Voskes \& Burton 2006, Levine et al.\ 2006);
our warp is somewhere between an L-warp and a S-warp rather than being a pure
S-warp. Moreover, the position of
the Sun  also affects our perspective of the northern and southern warp.
Also, there are some other parameters of the
disc (galactocentric distance of the Sun,
height of the Sun above the plane, flare parameters, scale length,
scale height at the Sun galactocentric distance, etc.) whose variations
affect the result. Furthermore, the thick-disc component
(Cabrera-Lavers et al.\ 2005), which was not included, could also produce
some small variations.

I am not going to explore  all the
set of parameters here but just the azimuth, $\phi _w$, of the warp; that
is, the galactocentric angle in which the warp has null amplitude.
It was checked that the galactic longitude of the maximum
$\rho (S)-1.2\rho (N)$ is more sensitive to the changes in $\phi _w$
than changes in other parameters.
For instance, changes in $\epsilon _W$ (with the corresponding change
in $C_W$ to preserve the amplitude at high $R$) mainly affect the
distance of the maximum overdensity rather than its galactic longitude;
also, other changes in the disc or flare  have less effect than changes
in $\phi _w$ with galactic longitude.
Figures\ \ref{Fig:simul}c--e show the density maps for angles
$\phi _w=0,+5^\circ $, $+10^\circ $ instead of the original value
$\phi _w=-5\pm 5^\circ $ given by L02.
As observed the variation in the position of the maximum is notable.
With $\phi _w=+5^\circ $ (2$\sigma $ from the value given by L02)
the angle of the maximum is 248$^\circ $.
Just to see that the position of the maximum is not so
sensitive to slight changes in the shape of the warp, we also performed
the calculations with 
Drimmel \& Spergel's (2001) formula for the warp amplitude
($25[R(kpc)-7]^2$ instead of $C_WR^{\epsilon _W}$
in expression (\ref{warp1}); the result was also 248$^\circ $
(Fig.\ \ref{Fig:simul}f). This galactic longitude is not
very far from the observed value of 240-244$^\circ $ by B06b, and its distance
of 5.4 kpc is somewhat less than the value $r=7.2$ kpc given by B06b.
The distance $r$ of the maximum could be changed if
the flare, warp and scalelength were rescaled, or if they showed a
different dependence on $R$ from those given by L02. 
For instance, a large scalelength for the disc and the flare, and a
lower exponent $\epsilon _W$ places the maximum farther away.

These plots may be compared in  
fig.\ 6 (top) of B06b. The similarity
is quite high (although not for $r$ larger than 9--10 kpc since in this case the
B06b red clump maps are highly contaminated by the dwarf population and 
 are very inaccurate).
Taking into account that we have only modified one parameter, the result
is not so bad. An angle of $\phi _w=+5^\circ $ is only 2$\sigma $
from the mean value in the fit of the average symmetric warp in the star 
counts. Other models of the average warp also give similar values: $\phi _W=
0$  (Freudenreich 1998; Drimmel \& Spergel 2001; Robin et
al. 2003) or +15$^\circ $ (Yusifov 2004, M06).

We must also consider
that the error of 5 degrees for $\phi _W$ in the analysis of L02 is 
merely statistical, but there might be
further systematic errors due, for instance, to errors in
the assumed luminosity function of stars, 
or deviations from equation (\ref{warp1}).
As M06 say, some uncertainties of the L02 model could also come
from gaps in the region around $l=270^\circ $ in the data used
for the fit. However, M06 are not correct when they claim that the L02 warp
is affected by contamination of dwarfs for red clump stars,
because L02 do not use red clump stars for the determination of the
parameter of the warp but only the total star counts.

As has been said, given the irregularities of the southern warp, 
it is quite possible that an extra shift in the azimuth 
in the southern with respect the 
average L02-warp might occur. Indeed, the analysis by Voskes \& Burton (2006)
gives precisely the best fit for $\phi _w=+5^\circ $ (warp maximum at
$\phi =95^\circ $) for their gas warp. The asymmetries in
the gas warp (similar to the stellar warp according to L02) 
are shown by Voskes \& Burton (2006) or Levine et al.\ (2006), 
which should serve to indicate that there is no unique large-scale 
symmetric warp model of the type eq. (\ref{warp1}); 
 therefore, we cannot extrapolate
the exact result at around $l\approx 240^\circ$ to the whole Galaxy.
Figure 11 of Levine et al.\ (2006) shows how the line of nodes defined
by $\phi _w$ varies depending on the galactocentric distance and from
north to south. Indeed, Levine et al.\ (2006) explain the asymmetries
with a warp dependence 
\begin{equation}
z_w(R,\phi )=W_0(R)+W_1(R)\sin [\phi -\phi_1(R)]
\label{warp2}\end{equation}\[
+W_2(R)
\sin [2\phi -\phi_2(R)]
.\]
The fact that $\phi _1\ne \phi _2$ [the difference is up to 12 degrees
according to Levine et al.\ (2006)] and depending on $R$, causes
the equivalent $\phi _w$ in expression (\ref{warp1})  not to be constant. 
The first term, the mode of $m=1$, is dominant for $R<13$ kpc, 
so eq.\ (\ref{warp1}) can be a relatively
good approximation, but if we aim to provide an accurate explanation
for all the deviations from it, perhaps eq.\ (\ref{warp1}) is insufficient. 
Here we are not making a fit of the parameters of the global warp ;
instead, we are showing that within the uncertainties the warp is 
compatible with B06b's plots.

M06 claim that {\it a global and regular warp signature
is traced to Galactocentric distances of at least $\sim 20$ kpc}, and 
the north/south asymmetry, apart from the fact that 
$\phi _W\ne 0$, is due to the chance location of the northern warp behind 
the Norma--Cygnus spiral arm. Perhaps this spiral arm is responsible
for the asymmetry, but data such as those illustrated in fig.\ 2 of Levine et
al.\ (2006) show something else: a lower amplitude for the southern warp
for $R>13$ kpc (see \S 2.2), and this is indeed necessary to explain
the distribution of sources along the line of sight, something which
was not successfully done by M06.

\

\subsection{Bump}
\label{.bump}

Another comment concerning B06b paper is their claim
that the bump in their fig.\ 1, in the region $238^\circ<l<244^\circ $,
$-11^\circ <b<-6^\circ $,
is due to the dwarf galaxy. In this case,
they compared it with the predictions of another warp + flare 
model (Robin et al.\ 2003) and failed to reproduce it. 
However, the bump could be explainable in terms of
the warp with an appropriate extrapolation for
values of $R>13$ kpc.
In \S 2.1, we saw that the extrapolation is not relevant for 
the position of the maximum overdensity. However, it is relevant
if we want to analyse the sources beyond 13 kpc.
Taking the southern warp equal to the northern warp is
inappropriate because there is an abundance of  data for the gas emission
of the Galaxy that show the asymmetry. 
At $R>13$ kpc, the $m=2$ mode becomes important and 
the extrapolation of the southern and northern warp are not equal.
Burton (1988) shows that the southern warp is approximately of constant
height between $1.6R_\odot $ and $2R_\odot $ (beyond $2R_\odot $
is unimportant for our analysis) instead of the monotonically
increasing northern warp. Something similar is observed in 
Voskes \& Burton (2006, fig.\ 16), and we know that the
gas warp and stellar warp are similar (L02, M06), so the
adopted approximate extrapolation for stars is justified.

\begin{figure*}
\vspace{1.5cm}
{\par\centering \resizebox*{12cm}{10cm}{\includegraphics{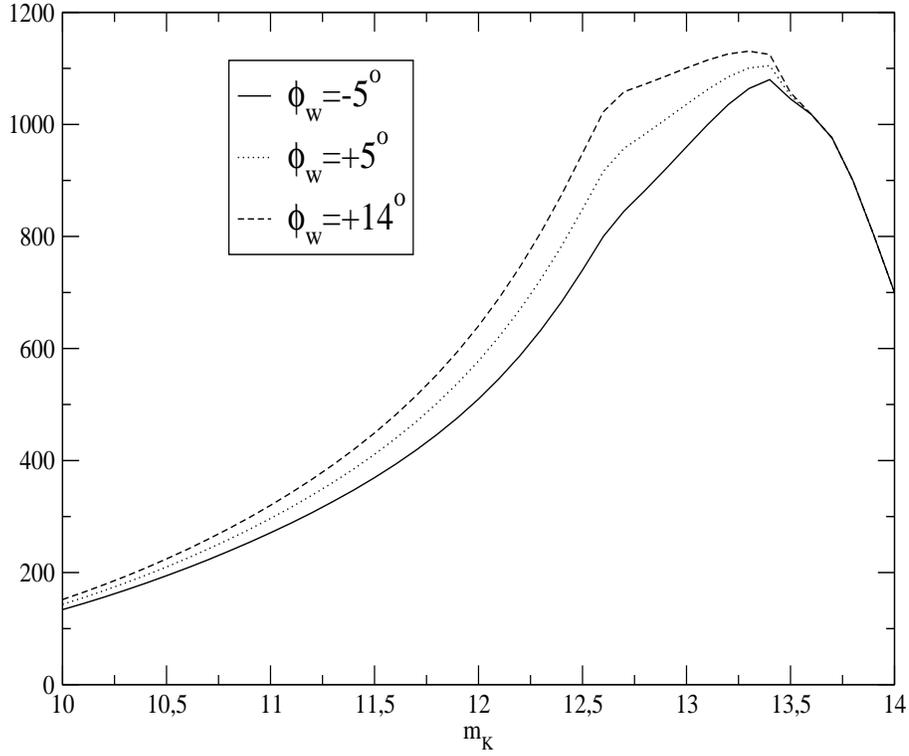}}\par}
\caption{Prediction of the warp model using L02 parameters for the counts
of the Fig. 1 in the paper by B06b ($238^\circ \le l\le 244^\circ$,
$-11^\circ \le b\le-6^\circ$) with extrapolation of constant height
in $13<R<16$ kpc and null beyond.}
\label{Fig:bump}
\end{figure*}

Compare Fig.\ \ref{Fig:bump} with fig.\ 1 of B06b.
In Fig.\ \ref{Fig:bump}, we see a bump with a maximum peak 
at $m_K=13.4$ (equivalent
to $R=15.4$ kpc), close to the peak at $m_K=13$--13.5 obtained
in B06b. Note, however,
that only the red clump giants are plotted in Fig.\ \ref{Fig:bump}, while
fig.\ 1 (right panel) of B06b includes all contaminants, especially dwarfs over $m_K=13.5$.
Indeed, the decrease in counts beyond $m_K=13$--13.5 is due to the end of the
warp at $R>$16 kpc, and, of course, the new increase in the counts around 
$m_K=13.6$--13.8 in fig.\ 1 of B06b would be due to the dwarf contamination.

\

\subsection{Colour--magnitude diagram and Canis Major depth}
\label{.depth}

Figure 7 of B06b gives a comparison of real and synthetic colour--magnitude 
diagrams. This plot is not easy to analyse by eye
because it includes CMa and foreground stars all together. 
And it depends on the  model of the warp used, so we 
should not say that since one
model of the warp does not fit, no model of a warp will fit. 
B06b see in their data a similarity with a 
Gaussian distribution of stars with an r.m.s. 
of 0.8 kpc. Could the warp produce such a colour--magnitude diagram?
The value of 0.8$\pm 0.3$ kpc that B06b take is from
Mart\'\i nez-Delgado et al.\ (2005a,b), who claim with their result
that the warp hypothesis is difficult to reconcile with
such a line-of-sight depth in the main sequence 
of a colour--magnitude diagram.

On the one hand, 
the Mart\'\i nez-Delgado et al. (2005a) paper contains errors:
i) the Gaussian fit of their fig.\ 3 gives $\sigma _{MS,total}=0.57$ mag;
once the contribution
of intrinsic broadening (0.19 mag) and latitude dispersion (0.29 mag) 
are subtracted, it would give $\sigma _{MS,CMa}=0.45$ mag 
(equivalent to 1.6 kpc at $d=7.9$ kpc instead of 0.8 kpc as they calculate);
ii) a Gaussian distribution of bins of constant magnitude does
not give a Gaussian distribution in the density distribution along
the line of sight (the second distribution is proportional 
to the first  multiplied by a factor
$1/d^3$), so we cannot translate  $\sigma $ in magnitudes into $\sigma $ 
in distances, that is, $\sigma =1.6$ kpc would be the r.m.s. of the function
$r^3\rho (r)$, not the r.m.s. of $\rho (r)$; 
iii) the assumed constant distribution for the underlying Milky Way stars 
(dotted line in fig.\ 3 of Mart\'\i nez-Delgado et al.\ 2005a) 
is not even a good first order approximation. 

\begin{figure}
\vspace{1cm}
{\par\centering \resizebox*{9cm}{7cm}{\includegraphics{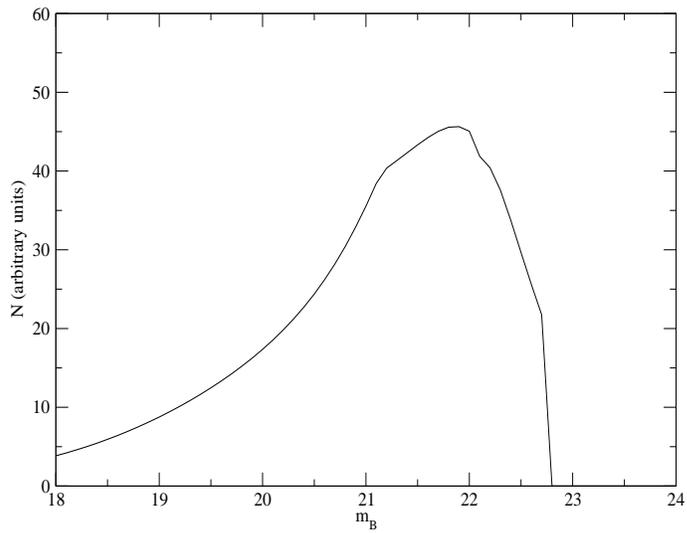}}\par}
\caption{Prediction of the warp model using L02's parameters,
except for $\phi _W$, which was changed to +5$^\circ $, for the counts
in fig.\ 3 in the paper by Mart\'inez-Delgado et al.\ (2005a) 
($l=240^\circ$, $b\le -8^\circ$)
with an extrapolation of constant height
for $13<R<16$ kpc and zero beyond.}
\label{Fig:david}
\end{figure}

On the other hand, if we forget the analysis 
by Mart\'\i nez-Delgado et al.\ (2005a) but  take as correct their data in their 
fig.\ 3 (a FWHM$\approx 1.6$ magnitudes in the distribution in bins of 
constant magnitude), we can compare them with our predictions of the warp and 
see that it is not so different with respect to the predictions of the
warp (see Fig. \ref{Fig:david}): 
2.2 mag (2.4 mag if we took into account the intrinsic
broadening and the latitude dispersion) instead of 1.6 mag.
The prediction of the L02 warp gives a somewhat broader distribution,
possibly because the radial dependence is not very accurate.

M06 claim instead that this structure is a spiral arm.
Again, I do not agree with the argument by M06
although I agree with their general conclusion that the warp can explain
the observed facts. No spiral arm is needed; it is just a question
of a wrong calculation of the thickness by Mart\'\i nez-Delgado et
al.\ (2005a) and a warp with appropriate extrapolation over $R>13$ kpc.

The population attributed by Mart\'\i nez-Delgado et al.\ (2005a) 
to be a 1--2 Gyr old population in the blue plume of intermediate-age
open clusters belonging to the CMa dwarf galaxy
is indeed a young stellar population ($\le 100$ Myr) of the Galactic
spiral arms in the background of open clusters, not placed in the putative
CMa galaxy (Carraro et al.\ 2005). The reply of B06b to Carraro et al.\
(2005) seems insufficient. Another paper by Bellazzini
et al.\ (2006a) claims that the metallicity of the core of CMa
is $-0.4\le [M/H]\le -0.7$, a relatively old population; however,
this is also within the expectation for the outer disc; for 
an $R=13.1$ kpc, $[Fe/H]\approx -0.57$ is expected for the Galactic disc
according to the metallicity distribution by Cameron (1985).

In conclusion, I do not see in the analysis of colour--magnitude
diagrams of the Canis Major region any conclusive proof that we are 
observing a population different from that of our own warped Galaxy.

\

\subsection{Velocity of the CMa stars}
The bimodal distribution
in the radial velocity of M-stars (Martin et al.\ 2004), 
presented as a proof that Canis Major is not the warp, 
reflects two kinds of origins for the sources: one 
was artificially produced by template issues resulting from a 
fluctuating line spread function asymmetry during the different observing
nights, as recognized by the authors in a later paper (Martin et al.\
2005); and the other peak can be reproduced by the Galactic
rotation (M06). In any case, even if M06 were wrong, the kinematics
of the warp is somewhat complex and unknown, so we cannot discard it. 
Another recent claim of measured motion of CMa perpendicular to the
disc (Dinescu et al.\ 2005; Mart\'\i nez-Delgado et al. 2005b) should
not be considered as inconsistent with the expected motion of the warp
because indeed we do not know very much about how the warp was formed
or its subsequent evolution, whether it is steady or still oscillating 
with respect to the plane---there is no unique scenario, and
there are at least four possible hypotheses of warp formation
(Castro-Rodr\'\i guez et al.\ 2002, Sect. 1), each one offering different predictions 
on its motion---or whether the northern and southern parts should have
similar kinematics (given the  north--south asymmetry, it is possible that we
cannot compare them).
Moreover, as M06 state, the stars selected
to measure the proper motions might be contamination not associated
with CMa, and {\it ``the expected warp signature can be and is compatible
with negative vertical velocity''.}

\

\section{Discussion}

The L02 model with the modified $\phi _w \approx +5^\circ$ fails slightly to 
reproduce the CMa feature but not by so much (and we must bear in mind
that the method of producing the maps  also involves certain errors, since the 
red clump stars used are contaminated by dwarfs, late-type giants and other
spectral types in different ratios depending on the line of sight). 
The bump of red clump stars with $m_K=13$--13.5
around $l=241^\circ $, $b=-8.5^\circ $  
or the depth of the Canis Major structure are also explainable in terms of 
the warp (with the appropriate extrapolation between 13 and 16 kpc of 
constant height, as observed). 
The blue plume in the colour--magnitude diagram is
explicable in term of the spiral arm population. The velocity distribution
of the stars cannot be a proof to exclude it is a warp.
The question now arises as to whether it is absolutely necessary to 
invoke the existence of a new dwarf galaxy to explain the red clump stars. 

The two options (warp or dwarf galaxy) are usually chosen depending
on the methodology of analysis. Those authors who prefer the dwarf galaxy
hypothesis assume a fixed model
of the warp and tend to think that any departure of this model is due
to the existence of the new galaxy. However, we must always bear in
mind that the predicted warp features depend on the parameters of
the disc, the  warp itself, the stellar population, the kinematics, etc.;
and all this knowledge is not so accurate as to allow a perfect
agreement with all the data, specially for the warp. Those authors
who prefer the opposite hypothesis claim that whatever you observe is the warp,
using an ad hoc model of it (as the case in the present
paper with a modified $\phi _W$). Perhaps none of the methodologies
is appropriate.
I am neither in favour nor against the dwarf galaxy hypothesis.
It is quite possible that CMa is a dwarf galaxy, but due to the proximity
of the warp feature, for which we do not have very accurate information,
it is difficult to disentangle both effects.

\section*{Acknowledgments}

Thanks are given to T. J. Mahoney (IAC-Tenerife) 
for proof-reading of this paper, 
and to the referee, S\'ebastien Picaud,
for useful comments that helped improve the quality of the paper.

\onecolumn

\end{document}